# Proportional Approval Method using Squared loads, Approval removal and Coin-flip approval transformation (PAMSAC) - a new system of proportional representation using approval voting

Toby Pereira

## Abstract

Several multi-winner systems that use approval voting have been developed but they each suffer from various problems. Six of these methods are discussed in this paper. They are Satisfaction Approval Voting, Minimax Approval Voting, Proportional Approval Voting, Monroe's Fully Proportional Representation, Chamberlin-Courant's Rule, and Ebert's method. They all fail at least one of Proportional Representation (PR), strong PR, monotonicity or positive support. However, the new method described in this paper – Proportional Approval Method using Squared loads, Approval removal and Coin-flip approval transformation (PAMSAC) – passes them all. PAMSAC uses the squared loads of Ebert's method, but removes non-beneficial approvals to restore monotonicity. It also uses the Coin-Flip Approval Transformation (CFAT), where voters are "split" into two for each candidate they approve, and where one half of this split voter approves and the other half does not approve each candidate approved on the ballot. This restores positive support, and also makes the method equivalent to the D'Hondt party-list method for party voting. PAMSAC reduces to simple approval voting in the single-winner case. A score voting version is described that also reduces to simple score voting in the single-winner case.

## 1. Introduction

Approval voting allows voters to approve (or vote for) as many or as few candidates as they like of those standing (Brams & Fishburn, 1978). It is a very simple system for voters to use, and escapes the problem of "overvoting" that can occur with block voting. Block voting limits voters to voting for no more candidates than there are positions available. As such, it requires voters to keep track of exactly how many votes they have cast and it can lead to accidentally spoilt ballots, particularly where there are many positions available.

Despite its simplicity, converting approval voting into a reasonable method of multi-winner proportional representation has proved problematic. But given the usefulness and simplicity of approval voting, particularly when it comes to electing committees, it is important to have a proportional version of it that can be relied on to give sensible results in all cases.

This topic has existed since at least the 1890s with the work of Thorvald N. Thiele and Lars Edvard Phragmén. However, progress seemed to dry up over the next 100 years or so, and it is only in recent years that further developments have been made. Because of this, the field is still under-developed, and it arguably lacks a single method that fulfils enough desirable criteria. PAMSAC is an attempt to fill that void.

This paper discusses six existing multi-winner approval methods with respect to four criteria, and finds that none are in compliance with all four. PAMSAC is then defined and shown to comply with all four criteria.

## 2. Existing Methods and Criteria

### 2.1 Methods under consideration

**Satisfaction Approval Voting** (Brams & Kilgour, 2011): A voter's satisfaction score is the proportion of their approved candidates that are elected. For example, if a voter approves two candidates and one of these is elected, their satisfaction score is $\frac{1}{2}$. The set of candidates that maximises total satisfaction is elected.

**Minimax Approval Voting** (Brams et al, 2007): The Hamming distance between two ballots (or between a potential result and a ballot) is the number of candidates that appear on one but not both ballots (or the number of candidates in one of the potential result and the ballot but not both). The proximity weight of a ballot is the number of voters that cast that particular ballot divided by the average Hamming distance from this ballot to all the voters' ballots. For a potential winning set of candidates, the weighted Hamming distance for a ballot is its Hamming distance from this potential result multiplied by its proximity weight. The winning set of candidates is the one that minimises the ballots' maximum weighted Hamming distance.

**Proportional Approval Voting**: While often attributed to Forest Simmons, this method was originally developed by Danish mathematician Thorvald N. Thiele in the 1890s (Janson, 2015). Voters have a satisfaction score based on the number of elected candidates that they have approved. If they have approved *n* elected candidates, then their satisfaction score is $1+\frac{1}{2}+\frac{1}{3}+...+\frac{1}{n}$. The set of candidates that maximises total satisfaction is elected. If each voter approves all the candidates from one party and no other candidate (party voting), this is equivalent to the D'Hondt party-list method. If the satisfaction scores are instead $1+\frac{1}{3}+\frac{1}{5}+...+\frac{1}{(2n+1)}$ then for party voting it becomes equivalent to the Sainte-Laguë party-list method.

**Fully Proportional Representation** (Monroe, 1995): Each voter is assigned to a candidate where each candidate has the same number of voters assigned to them. The winning candidate set is the one that allows the most voters to be assigned to a candidate that they approved. The method wasn't presented as specifically an approval voting method, but this is the approval version of it.

**Chamberlin-Courant's Rule** (Chamberlin & Courant, 1983): As Fully Proportional Representation except that each candidate need not have the same number of voters assigned to them. Candidates are given a weight in parliament based on the number of voters they represent. And as with Fully Proportional Representation, this is purely the approval version of the method.

**Ebert's method** (Bjarke Dahl Ebert [Note 1]): Each elected candidate has a "load" of 1 that is spread equally among their approvers (every elected candidate must be approved by at least one voter). For example, if an elected candidate is approved by 100 voters, each of these voters would have a load of $\frac{1}{100}$ from this candidate, which would be added to their loads from the other winning candidates. The winning set of candidates is the one that minimises the sum of the squared voter loads. For example, if a voter approves two elected candidates who each had 100 approvers, this voter would have a squared load of $\left(\frac{1}{100}+\frac{1}{100}\right)^2 = \frac{1}{2500}$. This would be added to the squared loads of every other voter. The concept of loads originally came from a sequential method developed by Swedish mathematician Lars Edvard Phragmén in the 1890s. [Note 2] Phragmén's method is not being considered in this paper because being a sequential method, it does not give a proper quality measure of each possible winning set of candidates that can be evaluated.

## 2.2 Criteria

There are several criteria that a proportional approval system would ideally fulfil. This paper considers four in particular that none of the above methods fulfil all of, but that the new method described in this paper (PAMSAC) does. They are proportional representation (PR), strong PR, monotonicity and positive support.

**Proportional Representation**: In its most basic form, if a system obeys PR, it means that for party or factional voting, each party or faction would win at least the number of seats equivalent to their proportion of the vote, rounded down to the nearest integer. We can call this a party or faction's minimal proportional allocation. Factional voting is the same as party voting except that while there are not explicit parties, voters approve candidates as if they are in parties. So for any two candidates, every voter must approve both or neither. Candidates that are always approved together are part of the same faction.

**Strong PR**: If one or more universally approved candidates are elected, and there is party voting for the remaining seats, then the breakdown of these remaining seats alone should be consistent with PR. For example:

3 to elect

$2n$ voters: *DEF*
$n$ voters: *G*

(Each letter refers to a candidate that the voters have approved.)

In this election, candidate *G* would be elected along with two of *DEF*: e.g. *DEG*.

6 to elect

$2n$ voters: *ABCDEF*
$n$ voters: *ABCG*

In this election, *ABCDEG* would be elected rather than *ABCDEF*. *ABCDEF* may look more proportional in a superficial sense because the $2n$ voters have 6 elected candidates and the $n$ voters have 3. However, they are not completely separate factions, and the universally approved candidates are not part of the "competition" between the two factions.

**Monotonicity**: If a candidate is given an additional approval, and no other ballot changes are made, then this will never cause this candidate to go from being elected to being unelected. Also, if candidate is *A* is approved by every voter that also approved candidate *B* and at least one other, then if candidate *B* is elected then so must candidate *A*. For example:

2 to elect

*n* voters: *AB*
*n* voters: *AC*

Monotonicity dictates that *AB* or *AC* must be elected, even if *BC* is the more balanced result and arguably the more truly proportional result.

**Positive Support**: We have a set of candidates, *X*, that would win an election with the currently submitted ballots. We then introduce a new block of voters for which set *X* is an exactly proportional result in a party or factional sense. For the same block of voters, the set *Y* is also proportional in the same sense. (For these purposes, *X* and *Y* are considered factionally distinct, meaning that a voter can be in a faction containing different voters for *X* as for *Y*, but there are consistent factions if you consider *X* and *Y* individually.) It is also the case that if we list the voters in descending order of the size of their faction for the result *X* alongside the equivalent list for *Y*, then at no position on the *X* list would the faction size be higher than at the corresponding position for *Y*, and at least one point on the list would the faction size for *Y* be higher than at the corresponding position for *X*.

If we clone this voter block enough times, then eventually result *Y* must defeat result *X*.

For example, *ABC* would be a proportional result for each of the following three ballot sets:

*n* voters: *A*
*n* voters: *B*
*n* voters: *C*

and

2*n* voters: *AB*
*n* voters: *C*

and

3*n* voters: *ABC*

We could then add further approvals to one of these ballot sets to make the result *DEF* proportional as well. For example, adding to the top ballot set:

*n* voters: *ADE*
*n* voters: *BDE*
*n* voters: *CF*

For the result *ABC*, *A*, *B* and *C* are the three factions. For the result *DEF*, *DE* and *F* are the two factions. Note that the result *ABC* could function as result *X* and *DEF* as result *Y* in the definition above. They are both proportional results for the above ballots, but *DEF* "dominates" *ABC* in the way described in the definition of positive support. If we divide the number of voters by *n* for simplicity, the descending list of faction sizes for *ABC* would be 1, 1, 1. For *DEF* it would be 2, 2, 1. The entry in the list for *DEF* is higher than for *ABC* in two places, and lower in none. We can say that *DEF* dominates *ABC* in terms of faction sizes.

The thinking behind this criterion is that merely being proportional is not enough. Some proportional results have more support than others and this should be reflected by the voting system, and even a certain amount of disproportionality is tolerable if there is enough extra support.

The simple "textbook" example of positive support is:

2 to elect

*xn* voters: *ABC*
*xn* voters: *ABD*
*n* voters: *C*
*n* voters: *D*

There must be a value of *x* where *AB* is elected over *CD*, despite *CD* representing every voter equally and *AB* failing to do so. Monotonicity would not require this.

As with the monotonicity criterion, pure proportionality is arguably sacrificed with the positive support criterion, but it would seem wrong to elect *CD* over the virtually unanimously supported *AB* when *x* gets very high and where *n* could even be 1.

## 2.3 Criterion failures of each method

**Satisfaction Approval Voting** fails PR with party voting. If a party fields too many candidates, then it will dilute the satisfaction of their voters too much and they won't get a proportional allocation of seats. For example:

6 to elect

*n* voters: *ABCDEF*
*n* voters: *GHIJ*

If *ABCGHI* are elected (as is proportionally correct), then the total satisfaction score is $\frac{3n}{6}+\frac{3n}{4}=1.25n$.

However, if *ABGHIJ* are elected (a disproportional result), then the total satisfaction score is $\frac{2n}{6}+\frac{4n}{4}\approx 1.33n$. This disproportional set of candidates would be elected. Therefore Satisfaction Approval Voting fails PR.

**Minimax Approval Voting** in its raw form fails PR. However, it can be modified to pass PR if voters are restricted to voting for exactly the same number of candidates as there are seats available, and ties are broken in favour of candidate sets that pass PR (Aziz et al, 2014). [Note 3] However, ties would be common given that a set of candidates is measured by the single ballot whose weighted Hamming distance to the candidate set is the greatest. It is not much of an achievement for a voting method to achieve PR in this way. Consider the following method: elect any set of candidates that allows the method to achieve PR. Also, every approved candidate is considered when determining the Hamming distances rather than just those in a set under evaluation. This means that the quality of a set of candidates is dependent on "irrelevant" candidates not in the set.

Finally, the restriction on voting that would need to be in place to achieve PR means that the system would not be true approval voting and is therefore outside the scope of this paper. As previously stated, with true approval voting, Minimax Approval Voting fails PR.

**Proportional Approval Voting** fails strong PR. For example:

6 to elect

$2n$ voters: *ABCDEF*
$n$ voters: *ABCG*

As explained above, a method passing the strong PR criterion would elect *ABCDEG* in this election. The two factions should be proportionally allocated seats if the universally approved candidates and their seats are ignored. However, Proportional Approval Voting effectively considers there to be two separate factions with $2n$ voters and $n$ voters, and awards them 6 and 3 seats respectively. It would therefore elect *ABCDEF*. On top of this, the addition of a third faction can cause Proportional Approval Voting to fail PR in a more basic sense. Partial agreement between two factions counts against both factions when competing against a third faction that has no split. For example:

20 to elect

$2n$ voters: *U1-U10, A1-A3*
$2n$ voters: *U1-U10, B1-B3*
$n$ voters: *C1-C6*

The ideal result here would seem to be *U1-U10, A1-A3, B1-B3* and *C1-C4*. The "*UA*" faction and the "*UB*" faction have $\frac{4}{5}$ of the voters between them, and the "*C*" faction has $\frac{1}{5}$. In terms of candidate numbers this result perfectly reflects that. However, Proportional Approval Voting would elect *U1-U10, A1-A2, B1-B2* and *C1-C6*.

This is because after all the "*U*" candidates are elected, Proportional Approval Voting effectively considers *UA* and *UB* to be separate factions, meaning that they should each have twice the number of candidates as the *C* faction (12 to 6 in this case), rather than four times the total of the *C* faction when considered together. The fact that they are not completely separate factions is not taken into account.

If the *UA* faction and the *UB* faction were either in full agreement with each other or in complete disagreement, then they would get 16 candidates between them and the *C* faction would get 4 candidates. But this partial agreement counts against the *UA* and *UB* factions, and works in favour of the *C* faction, leaving the *UA* and *UB* factions with 14 candidates between them and the *C* faction with 6. While this isn't a failure of PR in the simple way it has been defined in this paper, it would be a failure under a more nuanced definition. As a side note, it is a failure of "neutrality of partial agreement". This means that if two factions of voters both approve some but not all of the same candidates, it should not be to their advantage or disadvantage relative to other factions or voters. As well as the above example, the simple "textbook" example would be:

5 to elect

*n* voters: *U1, U2, A1*
*n* voters: *U1, U2, B1*
*n* voters: *C1, C2*
*n* voters: *D1, D2*

If we assume that *U1, U2, C1* and *D1* are elected, then there should be a four-way tie for the fifth seat between *A1, B1, C2* and *D2*. Proportional Approval Voting would elect *C2* or *D2*, as partial agreement counts to voters' disadvantage.

In any case, Proportional Approval Voting fails strong PR, which is enough for it to be discounted in this paper.

**Monroe's Fully Proportional Representation** and **Chamberlin-Courant's Rule** both fail monotonicity and positive support. They each assign a voter to a single candidate, so they are indifferent to a voter's opinion on all candidates other than "their" candidate. For example:

2 to elect

*n* voters: *ABC*
*n* voters: *ABD*

In this example, the methods would be indifferent between any pair of candidates. With the election of *AB, AC, AD, BC, BD* or *CD*, half the voters can be assigned to one candidate and half to the other, and these methods would declare them all to be proportional results. But not to elect both *A* and *B* would be a failure of monotonicity and positive support. A suitable tie-break method could restore monotonicity, but given the likelihood of ties with these methods, it seems that too much of the decision weight would be placed on the tie-break method, meaning that these methods are under-specified at present. In any case, no tie-breaker would restore positive support. For example:

2 to elect

*xn* voters: *ABC*
*xn* voters: *ABD*
*n* voters: *C*
*n* voters: *D*

The two methods would elect *C* and *D* for any value of *x*, meaning a failure of positive support.

**Ebert's method** fails both monotonicity and positive support. For example:

2 to elect

*n* voters: *AB*
*n* voters: *AC*

In this election, Ebert's method would elect *BC* because this result would give every voter an equal load, thereby minimising the sum of the squared voter loads. As previously said, this is arguably the most proportional result as it gives every voter the same level of representation. However, to elect *BC* over *AB* or *AC* would cause a failure of monotonicity. It also fails with the textbook example of

positive support:

2 to elect

*xn* voters: *ABC*
*xn* voters: *ABD*
*n* voters: *C*
*n* voters: *D*

Ebert's method would elect *CD* for any value of *x*, causing a failure of positive support.

We have now seen that none of the electoral methods so far described fulfil all of the criteria we desire. Specifically, each method fails at least one of PR, strong PR, monotonicity and positive support.

# 3. Proportional Approval Method using Squared loads, Approval removal and Coin-flip approval transformation (PAMSAC)

## 3.1 Ebert's method revisited

The new system being proposed here (PAMSAC) uses Ebert's squared loads as its base, so Ebert's method will now be considered in more detail.

By minimising the sum of the squared voter loads, Ebert's method works to minimise the difference between the voters' loads. For *c* candidates and *v* voters, maximum proportionality would be achieved if every voter had an equal total load of $\frac{c}{v}$.

If there was party voting, then equal loads would be achieved by each party winning the proportion of seats that corresponds to their proportion of the votes. If party X had *kx* voters (where *k* is a constant) and won *x* seats, then each voter of that party would have a load of $\frac{x}{kx} = \frac{1}{k}$. The differences between the numbers of voters for each party cancels out, leaving just the constant *k*, so if seats are awarded proportionally in this election, every voter would have a load of $\frac{1}{k}$.

For party voting, Ebert's method is equivalent to the Sainte-Laguë party-list method. This equivalence means that where there is an exact tie with the Sainte-Laguë method, there would be an exact tie with Ebert's method. This will now be demonstrated.

The Sainte-Laguë method works by sequentially awarding seats to parties based on their current weighted vote. Each time a party has a candidate elected, its vote weight is reduced. For *n* elected candidates, a party's vote weight becomes $\frac{1}{(2n+1)}$. For example, if a party has had three candidates elected so far, its effective number of votes would be the total number of votes cast for the party multiplied by $\frac{1}{(2 \times 3 + 1)} = \frac{1}{7}$.

We will take the case where party *A* has *ka* votes and party *B* has *kb* votes. Party *A* currently has $a - \frac{1}{2}$ seats and party B has $b - \frac{1}{2}$ seats.

Party A's effective vote count is $\dfrac{ka}{2 \times \left(a - \frac{1}{2}\right) + 1}$

$= \dfrac{ka}{(2a - 1 + 1)}$

$= \dfrac{ka}{2a}$

$= \dfrac{k}{2}$.

Since the *a* has cancelled out, the effective vote count would be the same for party B. There is therefore a tie for the next seat.

More generally, if party A has *ka* voters and party B has *kb* voters, then there would be a tie between these two results:

Party A: $a - \frac{1}{2}$ seats; Party B: $b + \frac{1}{2}$ seats

and

Party A: $a + \frac{1}{2}$ seats; Party B: $b - \frac{1}{2}$ seats

This will now be demonstrated for Ebert's method.

If party A wins $a - \frac{1}{2}$ seats and party B wins $b + \frac{1}{2}$ seats, then the sum of the squared voter loads is:

$ka \times \left(\dfrac{\left(a - \frac{1}{2}\right)}{ka}\right)^2 + kb \times \left(\dfrac{\left(b + \frac{1}{2}\right)}{kb}\right)^2$

$= \dfrac{(a+b) \times (4ab + 1)}{4abk}$.  Note 4

Note that the *a* and *b* are in equivalent positions here; the +/- $\frac{1}{2}$ has cancelled out so we can exchange them with no change in result. This means that if there is a tie between two parties for the next seat using the Sainte-Laguë method, there would also be a tie using Ebert's method. If one party would beat another party to the seat using Sainte-Laguë, then it also would using Ebert's method. We can compare any two parties against each other this way, so this is not limited to an election with just two parties standing. It follows that with party voting, the Sainte-Laguë party-list method is equivalent to Ebert's method used sequentially.

But Ebert's method is a global rather than a sequential method, so could it still yield different results? There are cases where Ebert's method used sequentially would give different results to Ebert's method used globally, but not with party voting. Because voters are perfectly partitioned into parties, if electing a candidate from party *A* increases the sum of squared voter loads by less than electing a candidate from party *B*, then this will remain the case regardless of which other parties subsequently have a candidate elected. There is no situation where we would want to reverse the sequential election of a party candidate when using a global outlook.

Therefore, for party voting, Ebert's method is equivalent to the Sainte-Laguë method.

Ebert's method also passes strong PR. In addition to party candidates, if there are also one or more elected candidates that are universally approved, then each voter would take an equal load from these candidates, equal to $\frac{u}{v}$ for *u* universally approved candidates and *v* voters. If the voters had equal total loads without the universally approved candidates, then this equality would remain with the addition of these candidates. Therefore, having universally approved candidates does not affect the correct party seat proportions, meaning that strong PR is obeyed. We get the same Sainte-Laguë tie when universally approved candidates are added. We can take the same example as before where party *A* wins $a-\frac{1}{2}$ seats and party *B* wins $b+\frac{1}{2}$ seats, but where there are also *u* universally approved candidates. The sum of squared voter loads would be:

$$ka \times \left(\frac{\left(a-\frac{1}{2}\right)}{ka}+\frac{u}{ka+kb}\right)^2 + kb \times \left(\frac{\left(b+\frac{1}{2}\right)}{kb}+\frac{u}{ka+kb}\right)^2.$$

The answer to this sum is fairly complex, but we know it should be equal to the sum of squared voter loads when we swap the party positions round so that party *A* wins $a+\frac{1}{2}$ seats and party *B* wins $b-\frac{1}{2}$ seats. And by subtracting one from the other, we can see that this is the case. [Note 5]

Another way of looking at this would be to sum the squares of the differences between each voter's load and the mean load instead of summing the squares of each voter's load. Minimising this would always give the same result as minimising the sum of the squared voter loads. This will now be demonstrated.

One formula for the variance of a variable *X* is $variance = mean[X^2] - (mean[X])^2$. In this case, *X* corresponds to the voters' loads. The mean of *X* is the mean load, which is $\frac{c}{v}$ for *c* elected candidates and *v* voters. The variance is the mean of the squared differences between the voters' loads and $\frac{c}{v}$. For a perfectly proportional result, this variance would be zero. For a given election, regardless of the result, $(mean[X])^2$ would always be the same. It would simply be $\left(\frac{c}{v}\right)^2$. Because of that, we can effectively ignore it in the equation. Whenever $mean[X^2]$ (the mean of the squared voter loads) is greater, the variance is greater. It also makes no difference whether we use total sum or mean when making comparisons. This means that whenever the sum of the squared voter loads is greater, the sum of the squared difference between each voter's load and $\frac{c}{v}$ is also greater. So they are equivalent measures when it comes to comparing candidate sets, which is the result we wanted.

But by using the difference between a voter's load and $\frac{c}{v}$ we can see more clearly that adding universally approved candidates into extra seats makes no difference to the result for the existing seats. Each extra seat means that $c$ goes up by 1 and $\frac{c}{v}$ goes up by $\frac{1}{v}$. An individual voter's load will also go up by $\frac{1}{v}$. This means that the difference between a voter's load and $\frac{c}{v}$ will remain unchanged with the addition of universally approved candidates and extra seats to accommodate them. Therefore, the winning result before the universally approved candidates are added will remain the winning result after they are added. And this is regardless of whether there is party voting or something more complex.

### 3.2 Approval Removal

As we have seen, Ebert's method in its raw form fails monotonicity. An approval for a candidate can cause them to fail to be elected. We have already seen this example:

2 to elect

*n* voters: *AB*
*n* voters: *AC*

Although candidate *A* is universally approved, a system that minimises the differences across voters' loads would elect *BC*. A way to avoid this result is to use "approval removal". For each possible winning set of candidates, we minimise the sum of the squared voter loads by removing voters' approvals as necessary (although no elected candidate can be reduced to zero approvals). The winning set would then be the one that had the minimum possible sum of squared voter loads using approval removal. Ties would then be broken by looking at the total number of approvals for each candidate set after approvals are removed, and as a second tie-break, the total number of approvals before any approvals have been removed.

In the above example, when considering the potential result *AB*, approvals could be removed to give the following ballots:

*n* voters: *B*
*n* voters: *AC*

*AB* would now match *BC*'s sum of squared voters loads and would then beat *BC* in the second tie-break (the first tie-break method would not break the tie). The equivalent could be done for *AC*, so *AB* and *AC* would be tied as the winning sets.

Although the first tie-break method did not break the tie in the above example, it is a necessary tool to retain strong PR. To reuse a previous example:

6 to elect

2*n* voters: *ABCDEF*
*n* voters: *ABCG*

With no approval removal, *ABCDEG* minimises the sum of the squared voter loads and beats *ABCDEF*. However, with approval removal, *ABCDEF* can match *ABCDEG*. Approvals can be removed to make the ballots look like this:

1.5*n* voters: *DEF*
0.5*n* voters: *ABC*
*n* voters: *ABCG*

Under *ABCDEF*, each voter would now have an equal load. This would therefore minimise the sum of the squared voter loads. The tie between sets *ABCDEG* and *ABCDEF* would now need to be broken. If we skipped straight to the second tie-breaker, then *ABCDEF* would win as it has more approvals before approvals are removed. However, the first tie-breaker looks at approvals after approvals have been removed. No approval removal was necessary for *ABCDEG*, so it has has $3\times 3n + 2\times 2n + 1\times n = 14n$ approvals. *ABCDEF* has $3\times 1.5n + 3\times 1.5n = 9n$ approvals. *ABCDEG* is therefore the winning set.

Approval removal creates a system that is guaranteed to be monotonic. If, for a candidate set being considered, an approval increases the sum of the squared voter loads, it is removed, so no approval can ever be harmful, creating at least weak monotonicity. And extra approvals would always count positively in tie-breaking scenarios.

Approval removal retains PR for party voting because for any result, the sum of squared voter loads for a party's voters is minimised by the party's voters all having equal loads. If a party's voters all approve every party candidate, then this will achieve equal loads, so no approval removal can reduce the sum of squared voter loads.

Strong PR also holds with approval removal. It is clear from the example above that if exact proportional apportionment is possible with party voting, then the existing seats would not be affected by the addition of extra seats filled by universally approved candidates. Every voter would still have an equal load, so the sum of squared voter loads would be at a minimum. Other potential results may achieve the same sum of squared voter loads, but would lose on the first tie-break. Where exact proportional apportionment is not possible among the "party seats", parties would still be guaranteed their minimal proportional allocation among these seats. By monotonicity, if a party's exact proportional allocation is higher than its minimum proportional allocation (as opposed to equal to it as in the above example – and it can never be lower), it would be guaranteed not to go below its minimum proportional allocation.

While a party is guaranteed their minimal proportional allocation among the non-universal seats, meaning that strong PR is obeyed, where exact proportional apportionment is not possible, the addition of universally approved candidates can cause the seat allocations to change. For example:

2 to elect

3*n* voters: *BC*
*n* voters: *D*

There would be a tie between all three possible results: *BC*, *BD* and *CD*. The sum of squared voter loads for *BC* would be $3n \times \left(\frac{2}{3n}\right)^2 = \frac{4}{3n}$. The sum of squared voter loads for *BD* and *CD* would be $3n \times \left(\frac{1}{3n}\right)^2 + n \times \left(\frac{1}{n}\right)^2 = \frac{4}{3n}$. This equivalence gives us a tie. As has already been demonstrated, no approval removal can reduce the sum of squared voter loads when there is party voting, so this tie remains with approval removal available. However, we can add a universally approved candidate to the above example as follows:

3 to elect

3*n* voters: *ABC*
*n* voters: *AD*

With no approval removal, we would still have a tie. The sum of squared voter loads for *ABC* would be $3n \times \left(\frac{1}{4n} + \frac{2}{3n}\right)^2 + n \times \left(\frac{1}{4n}\right)^2 = \frac{31}{12n} \approx \frac{2.58}{n}$. For *ABD* and *ACD* it would be $3n \times \left(\frac{1}{4n} + \frac{1}{3n}\right)^2 + n \times \left(\frac{1}{4n} + \frac{1}{n}\right)^2 = \frac{31}{12n} \approx \frac{2.58}{n}$. But using approval removal, we could change the ballots to the following:

$\frac{4}{3}n$ voters: *B*

$\frac{4}{3}n$ voters: *C*

$\frac{1}{3}n$ voters: *A*

*n* voters: *AD*

In this case, *A*, *B* and *C* each have a third of the vote – $\frac{4}{3}n$ – with no overlap, which minimises the sum of squared voter loads. It would be $4n \times \left(\frac{1}{\left(\frac{4}{3}n\right)}\right)^2 = \frac{9}{4n} = \frac{2.25}{n}$. But there is no way to achieve equal loads across all voters with the results *ABD* or *ACD* by approval removal. This is because candidate *D* is not approved by a third of the voters, which is the minimum required to be able to do this. So with approval removal, the candidate set *ABC* is elected. While approval removal retains strong PR, it does not retain full independence of universally approved candidates.

However, approval removal is not enough because it does nothing to help Ebert's method pass the positive support criterion. There must be a value of *x* above which *AB* beats *CD* in the following election:

2 to elect

*xn* voters: *ABC*
*xn* voters: *ABD*
*n* voters: *C*
*n* voters: *D*

The problem is that however high *x* gets, *CD* will always minimise the sum of squared voter loads, whereas *AB* will not. Every voter is equally represented by *CD*, whereas there is always some inequality under *AB*. Approval removal does nothing to address this. As discussed, *CD* is arguably a more proportional result, but it does not seem to be the correct result for high *x*.

## 3.3 Coin-Flip Approval Transformation (CFAT)

A way to fix the problem of positive support is to use the "Coin-Flip Approval" Transformation (CFAT). Despite its name [Note 6], it is deterministic. For every candidate that a voter approves, the voter is effectively split into two, where one half of the split voter approves the candidate and one half does not. For example, if a voter approves candidates *A*, *B* and *C*, their ballot is transformed as follows:

$\frac{1}{8}$ : *ABC*

$\frac{1}{8}$ : *AB*

$\frac{1}{8}$ : *AC*

$\frac{1}{8}$ : *BC*

$\frac{1}{8}$ : *A*

$\frac{1}{8}$ : *B*

$\frac{1}{8}$ : *C*

$\frac{1}{8}$ : -

This resolves the problem of positive support while retaining sufficient proportionality. However, while Ebert's method is equivalent to Sainte-Laguë party-list for party voting, with CFAT added it becomes equivalent to the D'Hondt method, which favours larger parties more. It effectively provides a "majoritarian shift". The equivalance with D'Hondt party-list will now be demonstrated.

Like the Sainte-Laguë method, the D'Hondt method works by sequentially awarding seats to parties based on their current weighted vote, but the weighting is done differently. For *n* currently-elected candidates, a party's vote weight becomes $\frac{1}{(n+1)}$. For example, if a party has had 3 candidates elected so far, its effective number of votes would be the total number of votes cast for the party divided by 4.

We will again take the case where party *A* has *ka* votes and party *B* has *kb* votes. Party *A* currently has *a* - 1 seats and party *B* has *b* - 1 seats.

Party *A*'s effective vote count is:
$$\frac{ka}{a-1+1}$$

$$= \frac{ka}{a}$$
$$= k.$$

As the *a* has cancelled out, the effective vote count would be the same for party *B*, so there is a tie for the next seat.

More generally, if party *A* has *ka* voters and party *B* has *kb* voters, then there would be a tie between these two results:

Party *A*: *a* - 1 seats; Party *B*: *b* seats

and

Party *A*: *a* seats; Party *B*: *b* - 1 seats

This will now be demonstrated for Ebert's method with CFAT.

For illustration purposes, the following is an example of how to calculate the load on a party's voters if they win 4 seats. The party has *v* voters and the number of seats they win (4 in this particular case) is *s*. *c* is the combination function. Below is the split in ballots of how many voters would approve 0, 1, 2, 3 or 4 of the 4 candidates after CFAT. In brackets is the calculation to give the voter split that approves the relevant number of candidates:

$\frac{v}{16}$ voters: 0 candidates ( $\frac{v}{16} = \frac{(sc\,0) \times v}{2^s}$ )

$\frac{4v}{16}$ voters: 1 candidate ( $\frac{4v}{16} = \frac{(sc\,1) \times v}{2^s}$ )

$\frac{6v}{16}$ voters: 2 candidates ( $\frac{6v}{16} = \frac{(sc\,2) \times v}{2^s}$ )

$\frac{4v}{16}$ voters: 3 candidates ( $\frac{4v}{16} = \frac{(sc\,3) \times v}{2^s}$ )

$\frac{v}{16}$ voters: 4 candidates ( $\frac{v}{16} = \frac{(sc\,4) \times v}{2^s}$ )

After CFAT is applied, $\frac{1}{2}v$ is the effective number of voters approving each of the party candidates, so a voter's load from a single candidate is $\frac{1}{\left(\frac{1}{2}v\right)} = \frac{2}{v}.$ So if a voter approves *t* of the *s* candidates, their total load is $\frac{2t}{v}$ and their squared load is $\left(\frac{2t}{v}\right)^2 = \frac{4t^2}{v^2}.$

The number of voters approving $t$ candidates would be $\dfrac{(sct) \times v}{2^s}$.

Therefore the sum of the squared loads of all the voters would be $\displaystyle\sum_{t=1}^{s}\left(\dfrac{4t^2}{v^2} \times \dfrac{(sct) \times v}{2^s}\right)$. We can cancel down a bit:

$$\sum_{t=1}^{s}\left(\dfrac{4t^2 \times (sct)}{2^s \times v}\right)$$

$$= \dfrac{s \times (s+1)}{v}. \quad \text{Note 7}$$

As previously, we will take the case where party $A$ has $ka$ voters and party $B$ has $kb$ voters. Party $A$ wins $a$ seats and party $B$ wins $b$ - 1 seats.

Using the above formula, the sum of the squared voter loads is:

$$\dfrac{a \times (a+1)}{ka} + \dfrac{(b-1) \times b}{kb}$$
$$= \dfrac{a+1}{k} + \dfrac{b-1}{k}$$
$$= \dfrac{a+1+b-1}{k}$$
$$= \dfrac{a+b}{k}.$$

Parties $A$ and $B$ are in equivalent positions here, so swapping round the minus one in the seat numbers between the two parties would give the same result. So as with the D'Hondt party-list method, if party $A$ has $ka$ voters and party $B$ has $kb$ voters, then there would be a tie between these two results:

Party $A$: $a$ - 1 seats; Party $B$: $b$ seats

and

Party $A$: $a$ seats; Party $B$: $b$ - 1 seats

This is the result we needed to show. And just as with the equivalence between Ebert's method in its raw form and the Sainte-Laguë party-list method, this equivalence is not affected by more than two parties standing or by non-sequential election. Therefore, for party voting, Ebert's method with CFAT is equivalent to the D'Hondt party-list method.

CFAT has no effect on strong PR. If there is a universally approved candidate, then the loads for the half of the transformed ballots that do not approve this candidate would be unaffected. The other half who do approve this candidate would be affected in exactly the same way as they would be without CFAT, meaning that strong PR is upheld.

As discussed, the shift towards D'Hondt is the result of a shift in a majoritarian direction at the expense of pure proportional representation. Sainte-Laguë is generally seen as being a purer form of PR than D'Hondt, although D'Hondt still passes the basic PR criterion. And it is this shift that solves the positive support problem. Positive support compliance will now be demonstrated.

We start with a set of ballots where set *X* wins, beating set *Y*. We now introduce a new block of ballots where set *X* and set *Y* are both proportional in the factional sense and where *Y* dominates *X* in terms of faction size. Because both results are proportional, we can say that under either of these results, then if a voter is in a faction with *v* voters, then they have elected *kv* candidates, where *k* is a positive constant. We also know from our D'Hondt-equivalance proof above that if a party has *v* voters and wins *s* seats, then after CFAT, the sum of squared voter loads for this party would be $\frac{s \times (s+1)}{v}$. In our current example with *v* voters and *kv* candidates, the sum of squared voter loads would be:

$$\frac{kv \times (kv+1)}{v}$$
$$= k \times (kv+1).$$

An individual voter's squared load would be:
$$\frac{k \times (kv+1)}{v}$$
$$= k^2 + \frac{k}{v}.$$

As *k* is a positive constant, the higher the value of *v*, the lower a voter's squared load. This means that the larger the faction that a voter is in, the lower their squared load.

Because result *Y* dominates result *X* in terms of faction sizes for this new voting block, we now know that the sum of squared voter loads for *Y* is less than it is for *X* within this block. Regardless of the margin of *X*'s victory over *Y* in terms of sum of squared voter load before the new voting block is introduced, if we clone this new block enough times, we can overturn a victory of any size. We can shrink the original ballots into being a negligible proportion of the total ballots. Therefore, combining Ebert's method with CFAT gives a method that obeys positive support. As already demonstrated, for party or factional voting, approval removal is of no consequence, so Ebert's method combined with approval removal and CFAT (PAMSAC in its complete form) obeys positive support.

To give the textbook example again:

2 to elect

*xn* voters: *ABC*
*xn* voters: *ABD*
*n* voters: *C*
*n* voters: *D*

Combining Ebert's method with CFAT means that *AB* is elected for *x* greater than 3, and *CD* for *x* less than 3. When *x* is 3, the sums of the squared voter loads are equal, meaning that *AB* wins on the first approval tie-break (no approvals are removed).

## 3.4 More precision

It seems sensible that two elections should return the same result if the number of voters that cast each possible ballot is multiplied by a constant. For example:

2 voters: *A*
2 voters: *B*
2 voters: *AB*

If these were some of the ballots cast, it may help candidates *A* and *B* to split up the two *AB* voters to give the following ballots:

3 voters: *A*
3 voters: *B*

However, an equivalent move is not possible with the following ballots:

1 voter: *A*
1 voter: *B*
1 voter: *AB*

Arguably, this sort of scenario would not make a difference to the result in a real-life election with many voters. However, it could in a very close election and, regardless of this, we would still want a system that is capable at "full power" of giving us the optimum result rather than an approximation. Because of this, it makes sense to be able to divide a voter into arbitrarily small fractions and remove the approval of any fraction. In the above example, we could split the *AB* voter into two, giving us:

1.5 voters: *A*
1.5 voters: *B*

But the approval removal must take place before CFAT. Removing approvals from transformed ballots can lead to invalid ballots. For example, if a voter approves candidates *A* and *B*, CFAT gives us these ballots:

$\frac{1}{4}$ : *AB*
$\frac{1}{4}$ : *A*
$\frac{1}{4}$ : *B*
$\frac{1}{4}$ : -

We cannot simply remove the *B* by the *A* to give us:

$\frac{1}{2}$ : *A*
$\frac{1}{4}$ : *B*

$\frac{1}{4}$ : -

This is because CFAT cannot produce this from any raw ballot. The ballot we end up with has to be something we can get from applying CFAT to a raw ballot.

## 3.5 Summary of method

To summarise PAMSAC – the measure of a set of candidates would be the minimum achievable sum of squared voter loads after approvals have been removed and the ballots have then been subjected to CFAT. The set of candidates that minimises this measure is the winning set. If there is a tie, the winning set is the one amongst the tied sets with the greatest number of approvals after approval removal. If there is still a tie, then it is broken by looking at the greatest number of approvals before approval removal. PAMSAC reduces to simple approval voting in the single-winner case, and to D'Hondt party-list for party voting.

## 3.6 Sainte-Laguë versus D'Hondt

Sainte-Laguë party-list is generally seen as a purer from of PR than D'Hondt party-list, and PAMSAC could be used without CFAT (so, PAMSA) to achieve a Sainte-Laguë version. However, this would be at the expense of positive support, which is probably an inescapable loss for a Sainte-Laguë version because of the majoritarian shift (away from true proportionality) required to pass this criterion. The Sainte-Laguë version would also rely heavily on tie-breaking to distinguish between results where there would be a clear winner under PAMSAC. For example:

2 to elect

$n$ voters: *ABC*
$n$ voters: *ABD*

PAMSAC would elect *AB* with no tie-break needed. The CFAT ballots for *AB* would be:

0.5$n$ voters: *AB*
0.5$n$ voters: *A*
0.5$n$ voters: *B*
0.5$n$ voters: -

The sum of squared loads would be:
$$\frac{1}{2}n \times \left(\frac{2}{n}\right)^2 + \frac{1}{2}n \times \left(\frac{1}{n}\right)^2 + \frac{1}{2}n \times \left(\frac{1}{n}\right)^2 = \frac{3}{n}.$$

The CFAT ballots for *CD* would be:

0.5n voters: *C*
0.5n voters: *D*
n voters: -

The sum of squared loads would be:

$$\frac{1}{2}n \times \left(\frac{1}{\left(\frac{1}{2}n\right)}\right)^2 + \frac{1}{2}n \times \left(\frac{1}{\left(\frac{1}{2}n\right)}\right)^2 = \frac{4}{n}.$$

As it has a lower sum of squared voter loads, *AB* would be elected. Without CFAT, *AB* and *CD* would equally minimise the sum of the squared voters loads, so *AB* would require a tie-break to win based on total approvals. While not fatal, it is not ideal. And it is this "knife-edge" win that causes a failure of positive support with a single voter that approves just *C* and *D*.

Instead of CFAT, which removes half of all approvals, it would be possible to instead remove a fraction less than a half. The smaller the fraction, the closer we would get to Sainte-Laguë proportionality, and any fraction greater than zero would be enough to pass the positive support criterion. But to get Sainte-Laguë equivalence exactly, the fraction would need to be reduced to zero, which would cause a loss of positive support.

### 3.7 Computability

PAMSAC is likely to be computationally expensive. Any proportional method that tests every possible winning set of candidates has the problem that for *k* candidates and *w* winners, there would be (*k c w*) sets to check, where *c* is the combination function. This is equal to $\frac{k!}{w! \times (k-w)!}$.

If many candidates are to be elected, then they could be elected sequentially. The first candidate to be elected would be the one with the greatest number of approvals. Then each following candidate to elect would be the one that gives us the best result (minimises the sum of the squared voter loads after approval removal and CFAT) given those already elected.

Finding the optimal approvals to remove would also be expensive. If there is a large number of voters, we could sacrifice the option of splitting voters in order to remove fractions of approvals, because it is unlikely to make a big difference. Alternatively, instead of being able to remove an arbitrary fraction of an approval, we could clone each voter a set number of times. This would retain some of the precision without as much computational expense, as only set fractions could be removed rather than any fraction between 0 and 1. Removing approvals could also be done in a "greedy" manner rather than checking every possibility separately.

For example, for a candidate set to be checked, each candidate's approvers could be listed in descending order of total load. Obviously voters have been "split", so each voter is no longer a single entity, but for these purposes we can just consider the fraction of each voter that has approved every candidate that the original voter approved on their ballot.

For *c* candidates, this gives us *c* lists. Now check the voter at the top of each list to see which approval removal would cause the greatest drop in the sum of the squared voter loads. Remove this approval and continue this process (re-sorting the lists each time) until no approval removal from a voter at the top of a list would cause a reduction in the sum of squared voter loads. If no approval removal from a voter at the top of a list can cause a reduction, then no approval removal from any voter can cause such a reduction because removing an amount of load from a voter with higher total load will always give a lower sum of squared loads than removing this same amount of load from a voter with a lower total load.

So at this point, even if it is not optimal, an equilibrium will have been reached where the sum of squared voter loads of a candidate set cannot be reduced by any candidate having an approval removed. It is thought to be unlikely that removing approvals greedily would differ greatly from a globally optimal solution, although this could be the topic of a future project.

## 3.8 Score voting conversion

It is possible to convert PAMSAC into a method that uses score rather than approval ballots. This can be done by first converting the score ballots into approval ballots and then applying the method as usual. This conversion is done before applying CFAT. There are several possible conversion methods, but my preferred one has become known as the "Kotze-Pereira Transformation" (KPT). [Note 8] For a maximum score of *m* and only integer scores allowed, each voter is split into *m* parts, numbered 1 to *m*. Part *p* (in the range 1 to *m*) approves every candidate that the voter has given a score of *p* or more.

For example, there is a maximum score of 5 and a voter has given scores of 5, 4 and 2 to candidates *A*, *B* and *C* respectively. Here is how the split would look (before CFAT):

Part 1: *ABC*
Part 2: *ABC*
Part 3: *AB*
Part 4: *AB*
Part 5: *A*

Each part is equivalent to a fifth of a voter in this case. KPT is preferred to other transformations because of its simplicity and also because it is scale invariant. That is to say that if all scores from all voters were multiplied by a positive constant, the result of any election would remain unchanged. PAMSAC with KPT also reduces to simple score voting in the single-winner case.

As with CFAT, approvals cannot simply be removed from a KPT transformed ballot. Scores have to be removed or reduced prior to any transformation to guarantee that the transformed ballot we end up with is something we can get by applying KPT followed by CFAT to a raw ballot.

A greedy removal system for score ballots would be more complex than for approval ballots. We would want to be able to reduce scores incrementally rather than remove them completely in one go. However, in the above example, reducing the score for candidate *A* by 1 point would never be beneficial because part 5 of the voter has only approved *A* and so will not be over-represented by having a high load. But it might still be beneficial to reduce the score for *A* by more than 1 point. Of course, we cannot start by removing part 1's approval of *A*, because that would give us an invalid ballot. Approvals have to be removed from the top down (from part 5 to part 1).

One possible method is as follows: As with the approval method, each candidate would have their own list of approvers in descending order of total load. But these approvers would just be the highest part of each voter that approved the candidate. In the above example, part 5 of this voter would be in *A*'s list, part 4 would be in *B*'s list and part 2 would be in *C*'s list.

Approvals would be removed (corresponding to a reduction in a voter's score by 1 point) as in the original method. Once this has been exhausted, we would look at reducing scores by 2 at a time. To do this, approvers of each candidate would be listed in order of total load of second-to-top part. In the above example, part 4 of this voter would be in *A*'s list, part 3 in *B*'s list and part 1 in *C*'s list.

Scores would be reduced by 2 at a time until there can be no further reduction in the sum of the squared voter loads by doing this. One caveat to this is that once a score has been reduced by 2, we would then have to check if it could be reduced by a further 1 before then carrying on checking the lists for scores that can be reduced by 2. All reductions by 1 would have been exhausted before the start of this "round", but this reduction by 2 would create a new possibility for a reduction by 1. If it could be reduced by a further 1, we would need to then check if it could be reduced by another 1 and so on.

Once reductions by 2 are exhausted, we would move on to reductions by 3 and so on until we reach an equilibrium where no score can be reduced or eliminated altogether to reduce the sum of the squared voter loads of the candidate set in question.

Because we are adding KPT to PAMSAC to create the score voting version, it can be known as "PAMSACK". That they have the same pronunciation is seen as a feature rather than a bug.

### 3.9 Participation criterion

One important criterion that has not been mentioned is participation. To pass participation, if a voter casts a ballot, then the number of elected candidates approved on this ballot must not ever be fewer than than the number of elected candidates approved on this ballot had they not cast the ballot. In other words, casting an honest ballot should not leave a voter worse off than if they did not vote at all. It has not been demonstrated whether PAMSAC passes participation, although no counter-examples have yet been found. Participation is an important enough criterion that any method claiming to be the "final word" on approval proportional representation would have to provably pass this criterion, or it would have to be proved that participation is incompatible with other criteria deemed more important. Proportional Approval Voting is known to pass participation.

## 4. Discussion and Conclusions

We have seen that PAMSAC, uniquely among the voting systems considered, passes PR, strong PR, monotonicity and positive support. PAMSAC or PAMSACK could be used for various elections, including potentially national parliamentary elections. PAMSAC would be particularly appropriate for electing committees that are normally elected by block voting. In block voting, voters are restricted to voting for no more candidates than there are positions available, and those elected are simply those with the most votes. The voting restriction is unnecessary and can lead to spoilt ballots, and electing based on total votes means that a well-organised majority faction can take all the positions rather than an arguably fairer proportional allocation.

PAMSACK in particular also has other uses outside elections: for example, film or book recommendations. If I wanted to make sure I'd seen all the "best" or most influential films, I could find a top 100 based on average scores and work my way down the list. However, there is likely to be a lot of similarity in this list, including many films that are liked by the same people because they have some of the same qualities. However, if the list was treated as a sequential election using PAMSACK, the list would be much more diverse, and would include films from more genres. In addition to this, a more interactive version would be where you could enter into the system which films you had already seen, and it would give you an "optimal" choice for your next film to watch.

This could be similarly useful for giving diverse search engine results. Arguably the most prominent use of a proportional score system to date is when Reweighted Range Voting (a sequential score version of Proportional Approval Voting, independently invented by Warren D. Smith) was used to select Oscar nominees. [Note 9]

To conclude, PAMSAC fills a gap that was present among multi-winner systems of approval voting. It passes PR, strong PR, monotonicity and positive support, and does so neatly by reducing to the D'Hondt party-list method for party voting, and to simple approval voting in the single-winner case. PAMSACK also reduces to simple score voting in the single-winner case.

## Acknowledgements

I would like to thank the following (in alphabetical order), who have all made contributions that have helped develop the ideas in this paper: Gabriel Bodeen, Jameson Quinn, Forest Simmons and Warren D. Smith.

## Notes

1: https://groups.yahoo.com/neo/groups/election-methods-list/conversations/messages/12468

2: https://groups.yahoo.com/neo/groups/election-methods-list/conversations/messages/10068, http://rangevoting.org/Phragmen.html

3: The proof in the paper was actually for a criterion the authors called "justified representation". However, the definition is close enough to PR as defined in this paper that the claims made here are unaffected.

4: http://www.wolframalpha.com/input/?i=k*a+*+((a+%E2%80%93+1%2F2)+%2F+(k*a))+%5E+2+%2B+k*b+*+((b+%2B+1%2F2)+%2F+(k*b))+%5E+2

5: http://www.wolframalpha.com/input/?i=k*a+*+(+(a-1%2F2)%2F(k*a)+%2B+u%2F(k*a%2Bk*b)+)%5E2+%2B+k*b+*+(+(b%2B1%2F2)%2F(k*b)+%2B+u%2F(k*a%2Bk*b)+)%5E2+-++k*b+*+(+(b-1%2F2)%2F(k*b)+%2B+u%2F(k*a%2Bk*b)+)%5E2+-+k*a+*+(+(a%2B1%2F2)%2F(k*a)+%2B+u%2F(k*a%2Bk*b)+)%5E2

6: Its name is due to Warren D. Smith.

7: http://www.wolframalpha.com/input/?i=The+sum+as+t+goes+from+1+to+s+of+4t%5E2+*+(s+combination+t)+%2F+(2%5Es+*+v)

8: Having independently discovered this transformation method, I found an earlier reference to it by an internet poster called "Kotze" - see http://www.revleft.com/vb/showpost.php?p=2030744&postcount=12 The name of the transformation is due to Forest Simmons. There was no vanity involved.

9: http://rangevoting.org/RRV.html, http://rangevoting.org/RRV.html#oscar